\def\spose#1{\hbox to 0pt{#1\hss}}
\def\approxlt{\mathrel{\spose{\lower 3pt\hbox{$\sim$}}
        \raise 2.0pt\hbox{$<$}}}
\def\approxgt{\mathrel{\spose{\lower 3pt\hbox{$\sim$}}
        \raise 2.0pt\hbox{$>$}}}
\def\approxpropto{\mathrel{\spose{\lower 3pt\hbox{$\sim$}}
        \raise 2.0pt\hbox{$\propto$}}}
\mathchardef\twiddle="2218

\def\multleft#1{\hbox to size{\vbox {\halign {\lft{##}\cr #1}}\hfill}\par}
\def\multright#1{\hbox to size{\vbox {\halign {\rt{##}\cr #1}}\hfill}\par}
\def\Mdot{\hbox{$\dot M$}}

\def\<{\thinspace}
\def\km{{\rm\thinspace km}}
\def\kpc{{\rm\thinspace kpc}}

\def\Mpc{{\rm\thinspace Mpc}}
\def\Msun{\hbox{$\rm\thinspace M_{\odot}$}}

\def\s{{\rm\thinspace s}}
\def\yr{{\rm\thinspace yr}}

\def\kmps{\hbox{$\km\s^{-1}\,$}}

\def\Msunpyr{\hbox{$\Msun\yr^{-1}\,$}}

\def\kmpspMpc{\hbox{$\kmps\Mpc^{-1}$}}

\long\def\symbolfootnote[#1]#2{\begingroup%
     \def\thefootnote{\fnsymbol{footnote}}\footnote[#1]{#2}\endgroup} 



\newcommand\beq{\begin{equation}}
\newcommand\eeq{\end{equation}}
\newcommand\beqa{\begin{eqnarray}}
\newcommand\eeqa{\end{eqnarray}}

\documentstyle[emulateapj,psfig,onecolfloat]{article}

\begin{document}

\twocolumn[

\title{The cosmological evolution of metal enrichment in quasar host galaxies}

\author{Tiziana Di Matteo\altaffilmark{1}, Rupert A.C. Croft\altaffilmark{2},
Volker Springel\altaffilmark{1}, Lars Hernquist\altaffilmark{3} }  

\altaffiltext{1}{Max-Planck-Institut f{\" u}r Astrophysik, Karl-Schwarzschild-Str.~1, 85740 Garching bei M{\" u}nchen, Germany}

\altaffiltext{2}{Carnegie-Mellon University, Dept. of Physics, 5000 Forbes Ave., Pittsburgh, PA 15213}  

\altaffiltext{3}{Harvard-Smithsonian Center for Astrophysics, 60
 Garden St., Cambridge, MA 02138} 

\begin{abstract}
  We study the gas metallicity of quasar hosts using cosmological
  hydrodynamic simulations of the $\Lambda$--cold dark matter model.
  Galaxy formation in the simulations is coupled with a prescription
  for black hole activity enabling us to study the evolution of the
  metal enrichment in quasar hosts and hence explore the relationship
  between star/spheroid formation and black hole growth/activity. In
  order to assess effects of numerical resolution, we compare
  simulations with different particle numbers and box sizes.  We find
  a steep radial metallicity gradient in quasar host galaxies, with
  gas metallicities close to solar values in the outer parts but
  becoming supersolar in the center. The hosts of the rare bright
  quasars at $z\sim 5-6$ have star formation rates of several hundred
  $\Msunpyr$ and halo masses of order $\sim 10^{12} \Msun$. Already at
  these redshifts they have supersolar ($Z/Z_{\sun} \sim 2-3$) central
  metallicities, with a mild dependence of metallicity on luminosity,
  consistent with observed trends. The mean value of metallicity is
  sensitive to the assumed quasar lifetime, providing a useful new
  probe of this parameter. We find that lifetimes from
  $10^{7}-4\times10^{7}\,{\rm yr}$ are favored by comparison to
  observational data. In both the models and observations, the rate of
  evolution of the mean quasar metallicity as a function of redshift
  is generally flat out to $z \sim 4-5$. Beyond the observed redshift
  range and out to redshift $z = 6-8$, we predict a slow decline of
  the mean central metallicity towards solar and slightly subsolar
  values ($Z/Z_{\sun} \sim 0.4 - 1$), as we approach the epoch of the
  first significant star formation activity.
\end{abstract}

 ]

\altaffiltext{1}{Max-Planck-Institute f{\" u}r Astrophysik, Karl-Schwarzschild-Str. 1, 85740 Garching bei M{\" u}nchen, Germany}

\altaffiltext{2}{Carnegie-Mellon University, Dept. of Physics, 5000 Forbes Ave., Pittsburgh, PA 15213}  

\altaffiltext{3}{Harvard-Smithsonian Center for Astrophysics, 60
 Garden St., Cambridge, MA 02138} 

\section{Introduction}

The presence of supermassive black holes in the centers of nearby
galaxies with a significant spheroidal component supports
arguments that there is a fundamental link between the assembly of
black holes and the formation of spheroids in galaxy halos. The
evidence indicates that the mass of the central black hole is
correlated with the bulge luminosity (e.g., Magorrian et al. 1998;
Kormendy \& Gebhardt 2001) and even more tightly with the velocity
dispersion of its host bulge (Tremaine et al. 2002; Ferrarese \&
Merritt 2000; Merritt \& Ferrarese 2001; Gebhardt et al. 2000),
implying that the process that leads to the formation of galactic
spheroids must be intimately linked to the growth of central
supermassive black holes with commensurate mass.

Quasars at high redshift, the black hole remnants of which we find in
galaxies today, provide us with a direct tool for investigating the
relation between black hole and spheroidal formation. There are
indications that high redshift quasar hosts are often strong sources
of sub-mm dust emission (Omont et al.~2001; Cox et al.~2002; Carilli
et al.~2000,~2002), suggesting that quasars were common in massive
galaxies at a time when they were undergoing vigorous star formation.
Metal abundance studies based on the gas of the broad line region
(BLR) of quasars (e.g. Hamann \& Ferland 1993; 1999; Hamann et
al. 2002; Dietrich et al. 1999; 2003a,b and references therein)
provide valuable information on the chemical enrichment of quasar host
galaxies. In particular, these studies have shown that the
interstellar medium of the quasar host galaxies has a metallicity
which is typically super-solar with a mean value $Z \sim 4-5\,
Z_{\sun}$, showing no evolution within the observed redshift range
$3.5 \le z \le 5$.  In the context of galaxy evolution, this must mean
that quasars mark the loci where massive galaxies are being assembled,
undergoing significant star formation and building the central black
holes.

Quasar metallicities offer us a potentially powerful probe of the
Universe at redshifts beyond those at which objects have been detected
so far.  As pointed out by Dietrich and coauthors, the star formation
timescales necessary to build up a reservoir of metal-rich gas by $z
\sim 4-5$ indicate the presence of massive star formation already at
$z\sim 6-10$. Other than the possible signature of reionization at
$z\sim 17-20$ seen by the WMAP satellite (Kogut et al.~2003),
information on the Universe at these epochs is difficult to
obtain. By
modeling the evolution of structure from these earliest times on it is,
however, becoming possible to make quantitative predictions for quasar
metallicities. The number density of rare luminous quasars is a
stringent test of models (Efstathiou and Rees 1988), and their
metallicities offer both a consistency test and a window on star
formation at an epoch when the quasars were forming.  Such
measurements are complimentary to those of the metallicity of gas in
the intergalactic medium from quasar absorption lines. The metallicity
in the extreme environments close to the centers of hosts is also
likely to be sensitive to events at earlier epochs, and can tell us
about the timescales for the assembly of material in the galaxies
themselves. The star formation rates required to produce metal-rich
gas are sensitive to the structure of dark matter on small scales (see
e.g. Yoshida et al. ~2003a,b) as well as the parameters which govern the
stellar mass function.

Several groups have discussed the link between cosmological evolution
of QSOs and the formation history of galaxies (see e.g.;~Monaco et
al.~2000; Kauffmann \& Haehnelt~2000; Ciotti \& van Albada~2001;
Wyithe \& Loeb~2002; Granato et al.~2003; Haiman, Ciotti \& Ostriker
2003 and references therein). Here we use cosmological hydrodynamical
simulations coupled with a prescription for black hole activity in
galaxies to follow the evolution of the metal enrichment in quasar
host galaxies and therefore explore the relation between star/spheroid
formation and black hole growth/activity. The simulations (Springel \&
Hernquist 2003a,b) include a novel prescription for star formation and
feedback processes in the interstellar medium that is able to
yield a numerically converged prediction for the full star
formation history.  In an earlier paper (Di Matteo et al.~2003), we
assumed that the black hole fueling rate is regulated by star
formation in the gas. We showed that this simple assumption can
explain the observed black hole mass and spheroid velocity dispersion
relation and the broad properties of the quasar luminosity functions
(for an assumed quasar lifetime). It also leads to a one-to-one
relation between black hole accretion rate density evolution and the
star formation history, which at high redshift implies that the quasar
phase and star formation rates both grow in response to the growth of
the halo mass function.

In this paper, we investigate how chemically enriched the quasar hosts
are, how the enrichment depends on the quasar luminosity and quasar
environment, and we look at the differences in the properties of
quasar host galaxies at high redshifts and the present day. In
particular, we examine whether the super-solar metallicities inferred
from observations can be explained within the framework of our
cosmological star formation model and its prescription for the cosmic
star formation rate.

The rest of the paper is organized as follows. In \S 2, we review the
basic features of our simulations and their associated prescription for
star formation. In \S 3, we describe our analysis and derive gas
density and metallicity profiles as a function of halo mass at
different redshifts, and we compare results from simulations of
different resolution. In \S 4, we describe our predictions for the gas
metallicity around quasars, and we discuss implications for the
properties of quasar hosts derived from the simulations up to
$z\sim8$. In \S 5, we compare the observationally inferred metallicity
with the evolution of gas metallicities around quasars from redshift
$z = 0$ to $z =8$ as a function of the quasar lifetime.  We discuss
our results and their implications in \S 6.
\begin{table*}
\begin{center}
\caption{\label{table_simul}
Simulations used in this study. }
\begin{tabular}{ccccccccc}
\tableline\tableline\\
Run  &  Boxsize & $N_{p}$ & $m_{\rm DM}$ & $m_{\rm gas}$ & $\epsilon$& $z_{\rm end}$ & wind & color\\
     &  $h^{-1}$Mpc &&  $h^{-1} \Msun$ &$h^{-1} \Msun$ &  $h^{-1}$\kpc & & &\\
\tableline\\
 Q5 & 10.00 & $2\times 324^3$& $2.12 \times 10^{6}$ & $3.26 \times10^{5} $ & 1.23& 2.75 & strong& {red}\\
 P4 & 10.00 & $2\times 216^3$& $7.16 \times 10^{6}$ & $1.1 \times 10^{6}$  & 1.85 & 2.75& weak & black\\
 D5 & 33.75 & $2\times 324^3$& $8.15\times 10^{7}$ &  $1.26\times 10^{7}$&  4.17& 1.00 & strong & blue\\
 G5 & 100.0 & $2\times 324^3$& $2.12 \times 10^{9}$ & $3.26\times 10^{8}$&  8.00& 0.00 & strong & black \\
 G6 & 100.0 & $2\times 486^3$& $6.28 \times 10^{8}$ & $9.66 \times 10^{7}$& 5.33& 0.00 & strong & pink \\
\tableline\\
\end{tabular}
\end{center}
\vspace{-1cm}
\tablenotetext{}{\small \\
%\tablecomments{*}
$N_p$ is the particle
number of dark matter and gas, $m_{\rm DM}$ and $m_{\rm gas}$ are the
masses of the dark matter and gas particles respectively, $\epsilon $
is the comoving gravitational softening length (a measure of the
spatial resolution) and $z_{\rm end}$ the ending redshift of the
simulation. The last column indicates the color which is used
in the figures to represent the results from the respective simulation.}
\end{table*}
\section{Simulations and analysis}
Throughout, we shall use a set of cosmological simulations for a
$\Lambda$CDM model, with $\Omega_{\Lambda}=0.7$, $\Omega_{\rm m}=0.3$,
baryon density $\Omega_{\rm b}=0.04$, a Hubble constant $H_{0} = 100\, h 
\kmpspMpc $ (with $h=0.7$) and a scale-invariant primordial power spectrum
with index $n=1$, normalized to the abundance of rich galaxy clusters
at the present day ($\sigma_{8} =0.9$).  Here, we briefly summarize the
main features of our simulation methodology and refer to Springel \&
Hernquist (2003a,b) for a more detailed description.

Besides self-gravity of baryons and collisionless dark matter, the
simulations follow hydrodynamical shocks, and include radiative
heating and cooling processes of a primordial mix of helium and
hydrogen, subject to a spatially uniform, time-dependent UV background
(see, e.g., Katz et al. 1996, Dav\'e et al. 1999).  The  dark matter
and gas are both represented computationally by particles.  In the
case of the gas, we use a smoothed particle hydrodynamics (SPH)
treatment (e.g., Springel et al. 2001) in its fully adaptive 
entropy formulation
(Springel \& Hernquist 2002) to mitigate problems with overcooling
(e.g.;~Croft et al. 2001) and to maintain
strict energy/entropy conservation (see, e.g. 
Hernquist 1993).

The model includes an effective sub-resolution treatment of star
formation and its regulation by supernova feedback in the dense
interstellar medium (ISM). In this model, the highly overdense ISM gas
is taken to be a two phase fluid consisting of a cold cloud component
in pressure equilibrium with a hot ambient phase.  Each gas particle
represents a statistical mixture of these phases.  Star formation is
assumed to occur in cold clouds that form by thermal instability (and
grow by radiative cooling) from a hot ambient medium, which is heated
by supernova explosions. These supernovae also evaporate clouds,
thereby establishing a tight self-regulation cycle for star formation
in the ISM.

The simulations keep track of metal enrichment and the dynamical
transport of metals by the motion of gas particles. Metals are
produced by stars which enrich the gas by supernova explosions.  The
mass of metals returned to the gas is $\Delta M_{Z} = y_{*}\Delta
M_{*}$, where $y_{*}=0.02$ is the yield, and $\Delta M_{*}$ is the
mass of newly formed stars. Assuming that metals are being
instantaneously mixed with the cold clouds and the ambient hot gas, the
metallicity $Z = M_{Z} /M_{g}$ of a star-forming gas particle
increases during one timestep $\Delta t$ in the simulations by \beq
\Delta Z = (1- \beta) y_* \frac{\rho_c}{\rho} \frac{\Delta t}{t_*},
\eeq where $\rho_c/\rho $ is the mass fraction of the gas in the cold phase,
$\beta$ is the mass fraction of stars that are short lived and
instantly turn into supernovae, and $t_*$ is the star formation
timescale, parameterized by \beq t_{*} = t_0^*
\left(\frac{\rho}{\rho_{\rm th}}\right)^{-1/2} \eeq in the model. Here the value
$t_0^*= 2.1$ Gyr is chosen to match the Kennicutt law (1998), and
$\rho_{\rm th}$ is the density threshold above which the multiphase
structure of the gas is followed and hence star formation can take
place.
\begin{figure*}[t]
\centerline{
\psfig{file=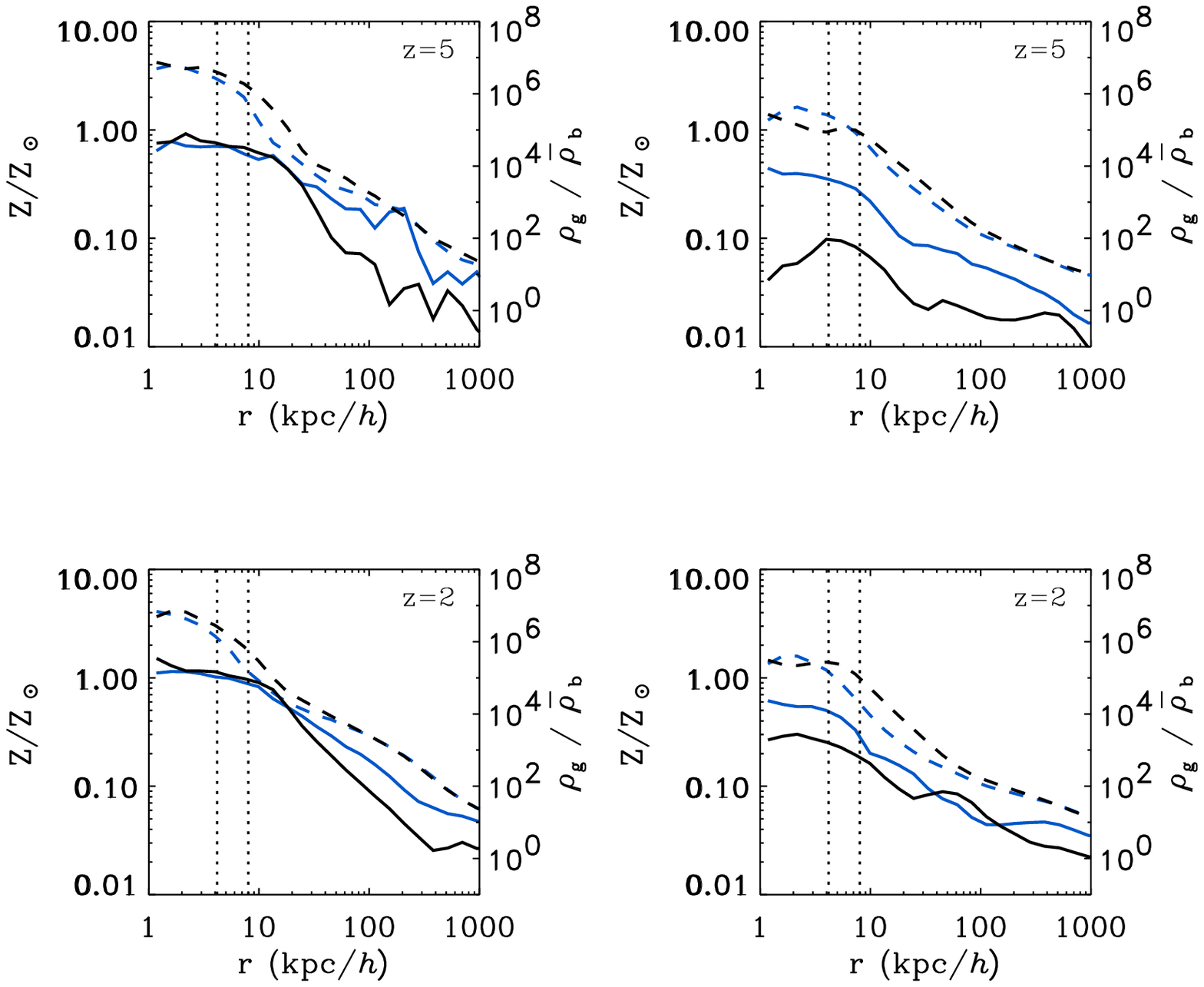,width=17.0truecm}}
\caption{Metallicity (left axis) and density profiles (right axis)
  shown with solid and dashed lines, respectively, of gas
  in halos of mass
  $M=2\times 10^{12} h^{-1} \Msun$ (left panels) and $M  =4\times 10^{10}
  h^{-1} \Msun$ (right panels) at two different redshifts. The black lines
  show the results from the G5 simulation and the blue lines from the
  D5 run.}
\label{fig_met_m}       
\end{figure*}

\begin{figure*}[t]
\centerline{
\hbox{
\psfig{file=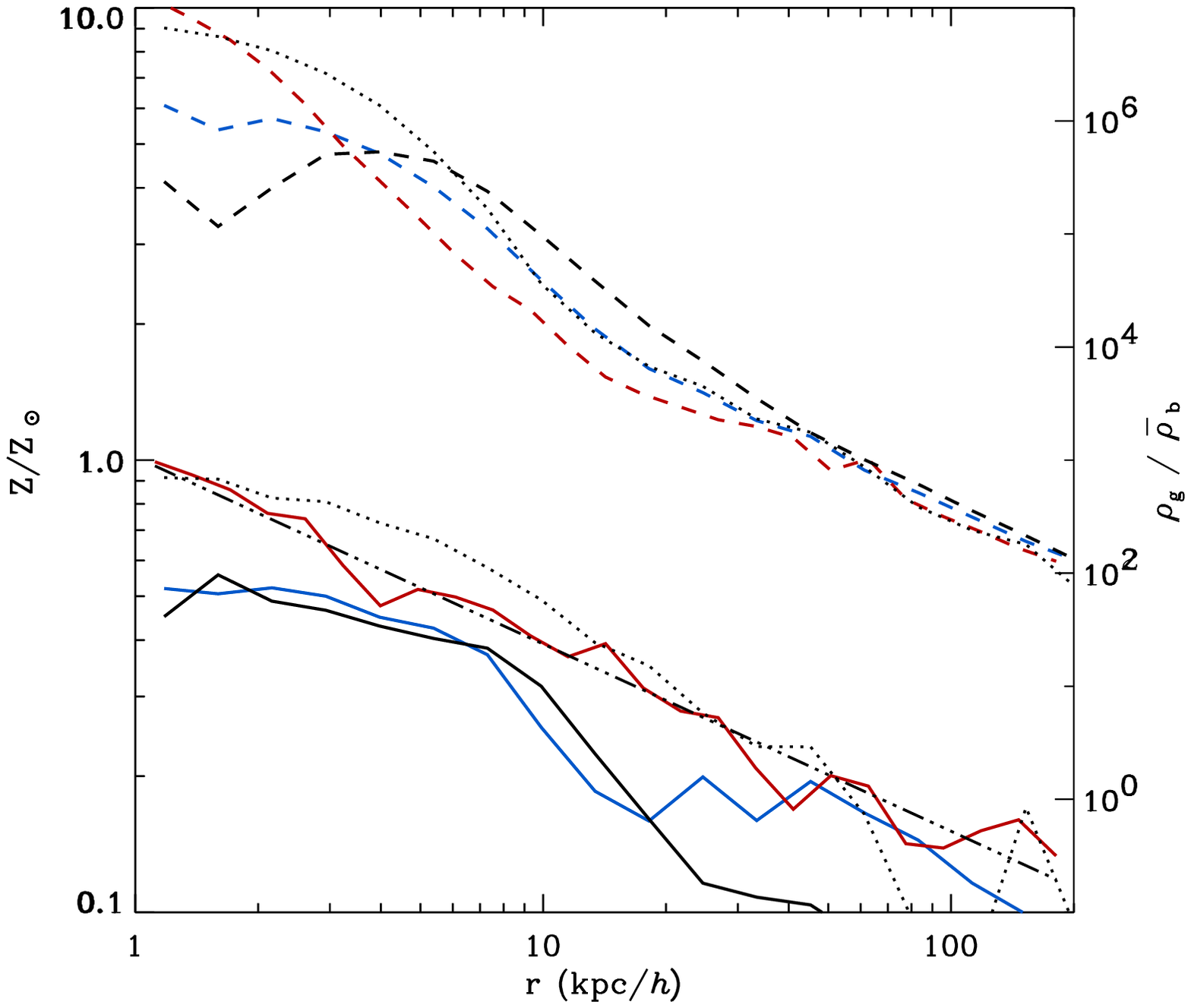,width=9.0truecm}
\psfig{file=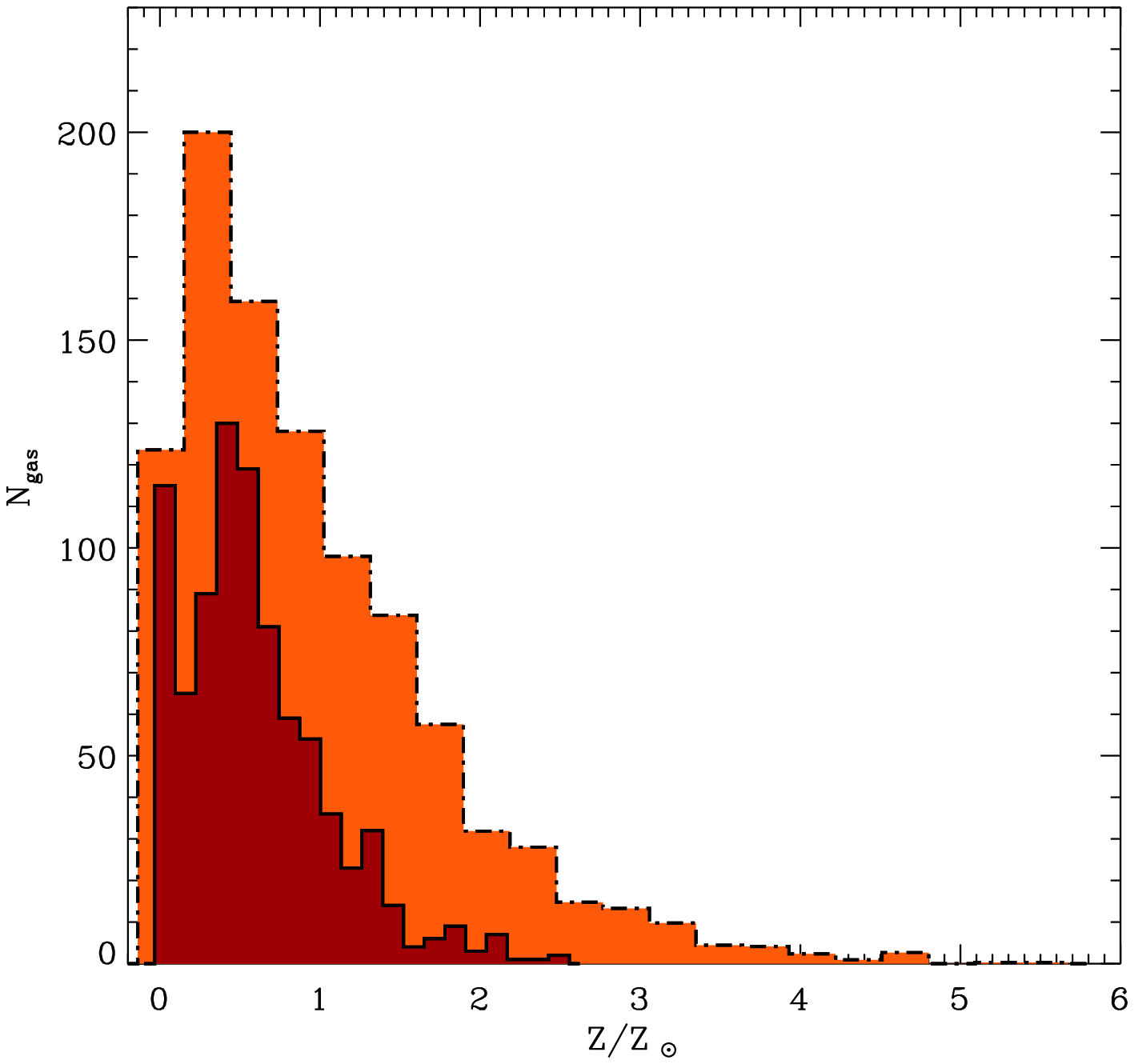,width=9truecm}
}}
\caption{Left panel: Metallicity (left axis) and density profiles (right axis)
  shown with solid and dashed lines, respectively, for halos of mass
  $M = 2\times 10^{11} h^{-1} \Msun$ (left panel) at redshift $z=3$. 
  The black
  lines show the results from the G5 simulation and the blue lines
  from the D5 run. The red lines are for the Q5 simulation, and the
  dotted lines are for the P4 run, which is a model with weaker
  galactic winds. Right panel: Histogram for the metallicity of
  particles within 2~\kpc\ in the two groups of mass $M=2\times
  10^{11} h^{-1} \Msun$ in the Q5 simulation. }
\label{fig_met_res}       
\end{figure*}

Finally, an important feature of the simulations is that they include
a phenomenological 
description of galactic winds. In this model, gas
particles are stochastically driven out of the dense star-forming
medium by assigning them extra momentum in random directions.  It is
assumed that the wind mass loss rate is proportional to the star
formation rate, $\dot{M}_{W} = \eta \dot{M}_{*}$, and that its kinetic
energy is comparable to the total available energy released by the
supernovae associated with star formation, $\frac{1}{2} \dot{M}_{W} v_{W}^2 =
\chi \epsilon_{\rm SN} \dot{M}_{*}$. The simulations we use adopt  a value
of $\eta =2$ and $\chi = 1$.  This parameterization leads to an initial
wind speed in the simulations equal to $484\,{\rm km\,s^{-1}}$.  (For
details, see e.g. Springel \& Hernquist 2003a; Aguirre et al. 2001a,b.)

In order to resolve the full history of cosmic star formation,
Springel \& Hernquist (2003b) simulated a range of cosmological
volumes with sizes ranging from $1\,h^{-1}$ Mpc to $100\,h^{-1}$ Mpc.  For
each box size, a {\em series} of simulations was performed where the
mass resolution was increased systematically in steps of $1.5^3$, and
the spatial resolution in steps of 1.5, allowing detailed convergence
studies.

In this work, we show results mostly from the 100 $h^{-1} \Mpc$
(runs G5 and G6), 33.75 $h^{-1} \Mpc$ (D5 run), and $10
h^{-1} \Mpc$ (Q5 and P4 runs) boxes within the G-, D-, and Q/P-series,
respectively (runs of the same boxsize are designated with the same
letter, with an additional number specifying the resolution).  
We note that the G6 run is a new higher resolution version of the
largest simulations described by Springel \& Hernquist (2003b).
The use
of these multiple runs is important for studying the effects of
numerical resolution. The Q/P-series can also be used to study the
effects of wind strength. The fundamental properties of 
the simulations (particle number, mass resolution etc.)
are summarized in Table~\ref{table_simul}. All length units
that we quote throughout the paper are comoving,
implying that at high $z$ the 
spatial resolution of the simulations in physical units is better than 
at low $z$.

In this work, it is important to analyze the larger boxes, as these
allow us to study the rare large massive objects that host the quasars
up to high redshifts and their evolution down to small redshifts. On
the other hand, the smaller box of the Q-series allows us to assess
the effects of numerical convergence and to study the physical
properties of halos down to small distances from their centers.

\section{Metallicity profiles versus halo mass}

We identify groups of gas, stars and dark matter in the simulations by
applying a conventional friends-of-friends algorithm to the dark
matter, gas, and star particles. The specific details of the grouping
method are described in Di Matteo et al.~(2003). The center of each
object is defined to be at the position of the gas particle with the
highest density. We find indistinguishable results if we consider
instead the gravitationally most bound particle. We then compute the
radially averaged profiles of density and metallicity about these
centers.

In Figure~\ref{fig_met_m}, we show the mean mass weighted
 metallicity profiles of gas (solid
lines) as a function of radius for dark matter halos of mass $M=2
\times 10^{12} h^{-1} \Msun$ (left panels) and $M=4 \times 10^{10} h^{-1}
 \Msun$ (right panels) at $z=2$ and $z=5$ (top and bottom panels,
respectively). Figure~\ref{fig_met_m} also shows the mean density
profiles normalized to the mean baryon density, $\bar{\rho_b}$.  The
metallicity and density profiles are calculated from both the G5 and
D5 runs and are shown by the black and blue lines respectively. The
vertical dotted lines represent the comoving gravitational softening
lengths of the G5 and D5 runs, which are $8 h^{-1} \kpc$ and $ 4.17
h^{-1} \kpc $, respectively (see also Table~\ref{table_simul}).  The
comparison of metallicity and density profiles obtained from the G5
and D5 runs (black and blue lines respectively) is particularly
instructive as we examine the effects of mass resolution on these
physical quantities. Understanding the variation of these effects
with radius will be important for our discussions in the rest of this
paper.

For the large mass objects (left panel; $M=2\times 10^{12}h^{-1}\Msun$),
the metallicity profiles are characterized by a flat compact core
extending to $r \sim 10\, h^{-1} \kpc$, where $Z/Z_{\sun} \sim
1$. Outside this core, $Z/Z_{\sun}$ declines more significantly in the
G5 run profiles than in the D5 run. Although in the inner regions the
two simulations lead to similar values for the metallicity, we should be
cautious in interpreting this, because the core component in the
G5 run, and for some of the objects in the D5 run, appears around or
within the simulations' respective gravitational softening lengths. The
presence of core components in the metallicity profiles may thus be
due to lack of sufficient spatial resolution.

This last point seems to be reinforced by an examination of the
density profiles. The mean gas density profiles measured in the G5 and
D5 runs show fairly good agreement in the outer regions (for
$\rho/\bar{\rho}_b
\approxlt 10^{3-4}$) but the G5 profiles imply typically higher densities
in the inner regions and again a flat density core within the
resolution limits. Compared to the D5 results, those for the
G5 run have enhanced densities in the core and a suppressed
metallicity in the outer regions. This seems to indicate a lack of fully 
resolved star formation and efficient transport of metals by its 
associated feedback processes, at least with respect to the D5
simulation.

For the smaller mass objects ($M=4 \times 10^{10} h^{-1}
\Msun$), the metallicity profiles in the G5 and D5 runs
show larger discrepancies, as expected. In particular, the central
metallicities measured in the G5 run are systematically lower than in
the D5 simulation, 
where the metallicity profiles imply values of $Z/Z_{\sun} \sim
0.4-0.8$. The density profiles show overall agreement in the outer
regions (at overdensities $ \rho/\bar{\rho}_b \approxlt 10^{2}$) but
the G5 profiles predict higher density values for distances less than
few tens of \kpc, as a result of underresolved star formation and
feedback in G5.

Although we expect the quasar hosts to reside in large mass halos,
such as those plotted on the left in Figure~1 ($M\sim 10^{12} h^{-1} \Msun$,
see also \S4.2), we would also like to understand the distribution of
metals in the central regions of less massive groups, which the D5 and
G5 runs are unable to resolve well. To do this, we turn to the Q5
simulation, which has much higher mass and force resolution
(comoving gravitational softening length equal to $1.2\,h^{-1} \kpc$). In
Figure~2, we plot the metallicity (and density) distribution for
objects in the Q5 simulation at $z=3$ (red lines). Because of the
relatively small box ($10\,h^{-1} \Mpc$ on a side) of the Q-series,
the largest objects have masses of order $M\sim 10^{11} h^{-1}\Msun$. We
also show a direct comparison with the profiles of objects of the same
mass measured in the D5 and G5 runs (blue and black lines,
respectively). Finally, we include results from the P4 simulation
(dotted line), which used the same size box as the Q-series but
included a weaker model for winds, and a slightly lower resolution
($N=2\times 216^3$).
\begin{figure}[b]
\centerline{
\psfig{file=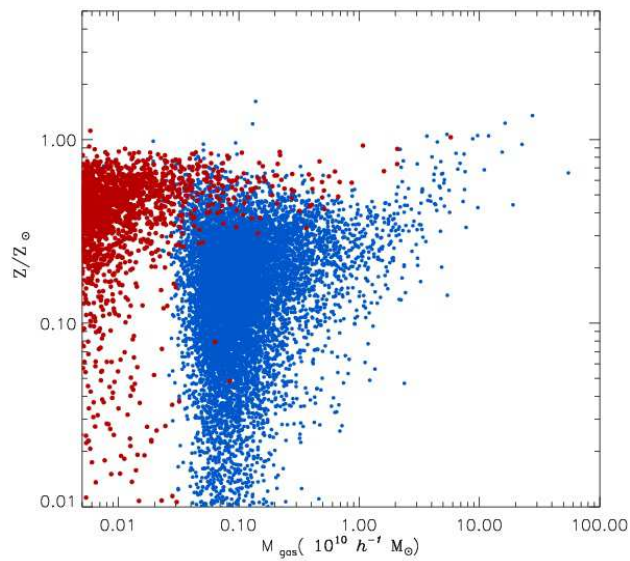,width=9.0truecm}
}
\caption{Central values for the metallicities measured in the Q5 (red) and
D5 (blue) simulations, as a function of 
the group gas mass in units of $10^{10} h^{-1} \Msun$. 
The D5 run tends to underestimate the central
metallicity values.}
\label{fig_met_mgas}       
\end{figure}

All the lines from the different simulations in Figure~2 represent
averages over 2-3 objects with similar mass (and are therefore
somewhat noisier than the profiles in Figure~1, which represented
averages over a larger number of objects). The most significant
feature to note in the comparison of the D5 and G5 results with the Q5
run is that while the metallicity profiles in the former simulations
still show a core-like component in the central regions, the profile
derived from the Q5 run continues to rise (the dash-dotted line
represents a fit to the Q5 profile with a power law, in $\log$, of
slope $\sim 0.4$) implying metallicity values in the center that are
at least a factor $\sim 2$ higher than in the D5 and G5 runs.

In accord with Figure 1, the gas density profiles in Figure 2 show
lower densities with improved resolution and fairly steep power law
behavior of the density and metallicity profiles extending to the very
central regions. Again this effect can be attributed to the improved
numerical resolution in the Q5 simulation, which allows us to resolve
higher gas densities and hence gas cooling, star formation and its
associated feedback processes over a much larger range
in scales.  Finally the
dotted lines in Figure~2 show the metallicity and density profiles
obtained from the P4 run. This simulation was run with a weaker wind
than the other simulations, with only half as much supernova energy
deposited as kinetic energy of the wind. Comparing with the other
profiles gives us some idea of the effects of the feedback by galactic
winds. Within P4, we see an excess of metals and gas density in the
center with respect to the other simulations. This is because the gas
in the P4 simulation can cool and form stars more efficiently, without
being blown away by winds as easily.

The right panel of Figure~2 shows a histogram of the metallicity
values of the gas particles within $r \le 2 h^{-1}\kpc$ for two groups in
the Q5 simulation with $M \sim 10^{11} h^{-1}\Msun$ (the profiles of which
are shown in the left panel).  This representation shows that the
majority of the gas in these central regions has metallicity around
solar (or below) but that there can be a relatively small fraction of
gas with significantly super-solar metallicities in the inner regions.

Finally, in Figure~\ref{fig_met_mgas} we show the values of the
density weighted metallicity within the central $2 h^{-1}\kpc$ of all the
objects in the Q5 run (red points) and those measured in the
objects of the D5 run within the central $5 h^{-1}\kpc$ (blue points).
The central metallicity values in the D5 run seem to be offset from
those measured in Q5 and are typically lower by a factor $\sim 2$.
Unfortunately, it is hard to assess whether this offset persists for
the large mass objects in D5 which cannot be probed with Q5.  However,
by comparing the profiles in Figure~\ref{fig_met_m} and
Figure~\ref{fig_met_res} we see no significant change in any of the
general trends described above, and hence expect that the same offset
in the central metallicity measurements is likely to persist up to the
largest masses. In other words, because of the significant gradient in
the metallicity profiles we have found in groups, we are not able to
obtain quantitatively accurate measurements of the circumnuclear
metallicity in the D5 or G5 simulations, because we are unable to
resolve the relevant scales below the gravitational softening lengths
of these simulations.  Also note that the large dispersion in the
central metallicity values in Figure~\ref{fig_met_mgas}, as measured
in the D5 run at around $M_{\rm gas} \sim 0.1\times10^{10}
h^{-1}\Msun$, is purely an
artifact of insufficient resolution. This is directly demonstrated by
comparing with the Q5 measurements.

In general, from the study of the metallicity profiles and the
resolution studies discussed in this section, we can conclude that
even within our model of quiescent star formation there appears to be
a fairly strong metallicity gradient in massive galaxies. Regions of
high density in the center of groups undergo strong star formation and
enrichment. As a consequence, metallicities up to solar and super-solar
values can easily be reached, at least in some fraction of the central
gas.  These results will be important for our discussions in the next
section, and later when we compare our model with several recent
measurements of circumnuclear gas in the high-$z$ QSOs.

\section{Quasar metallicities} 

\subsection{Quasar model}
So far, we have examined the properties of friends-of-friends selected
groups of gas and dark matter. In order to select quasar hosts from
these and determine the quasar properties from the simulations (in
which black holes have not been self-consistently included) we make
use of the model developed by Di Matteo et al.~(2003). Here we briefly
summarize the main characteristics of this scheme and refer the reader
to the above paper for details.  The model starts from
the hypothesis that black holes grow and shine by gas accretion, the
supply of which is regulated by the interplay with star formation in
spheroids. The black hole mass growth saturates in response to star
formation and its associated feedback processes. This results in 
black holes masses
that are related to the velocity dispersion of their
host spheroids in a manner which is consistent with the observed
$M-\sigma$ relation of Ferrarese \& Merritt (2000) and Gebhardt et
al. (2000). In this model, the QSO luminosity function evolves as a
result of the declining amount of fuel available for accretion (and
star formation), assuming a given (redshift independent) quasar
lifetime and duty cycle.  In Di Matteo et al.~(2003) we also showed
that within the context of this simple model the total
black hole accretion rate (BHAR) density very closely tracks the
star formation rate (SFR) density, as expected if black hole growth
and fueling are fundamentally linked to the assembly of the spheroids
and their star formation rate, respectively.

All galaxies are assumed to undergo active phases with a duty cycle
given simply by $f_{Q} =t_{Q}/t_{H(z)}$, where $t_{Q}$ is the quasar
lifetime and is the only free parameter of the model.  In order to
estimate the luminosity of each quasar, we simply assume that all the
gas in the galaxy is in principle available for accretion on this
timescale.  The bolometric luminosity owing to accretion onto the
central black hole at redshift $z$ is then given by
\beq 
L = \eta \Mdot(z) c^2
\sim \eta \frac {\Delta M_{\rm gas}}{\Delta t} c^2 \sim \eta f M_{\rm
  gas}(z) c^2/ t_{Q},
\label{eqn:Lacc}
\eeq where we have taken $\Delta M_{\rm gas} (z) \sim f M_{\rm gas}
(z)$, assuming that the average mass accreted is a constant fraction
$f$ of the total gas mass (see Di Matteo et al. 2003). We adopt the
standard value for accretion radiative efficiency of $\eta =
10\%$. With this model and $t_{Q} \sim 2-4 \times 10^7 \yr$, we were
able to reproduce the observed quasar luminosity function reasonably
well, particularly for $z \approxgt 2$. A general feature of this
model is that the black hole accretion rate history is found to
closely follow the cosmic star formation rate density, therefore
establishing a direct relation between black hole activity and star
formation. A consequence of this is that at high redshifts we expect
the evolution of black hole activity to be mostly driven by the
gravitational growth of structure, as this is the case for the star
formation rate density (Hernquist \& Springel 2003) relatively
independent of the details of the gas dynamics.
\begin{figure}
\centerline{
\psfig{file=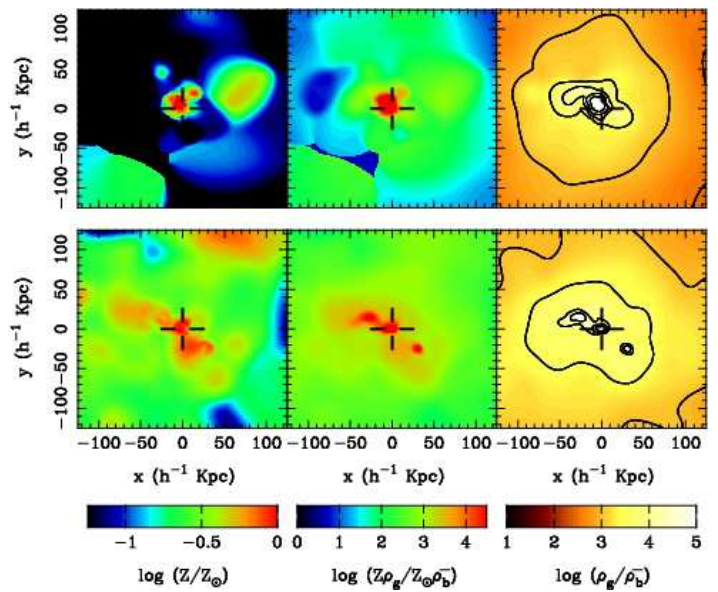,width=9.3truecm,angle=270}
}
\caption{Two quasar hosts in the G5 simulation, one at $z=5$ (top row) and
  one at $z=2$ (bottom row). For each host, we show from left to right
  the metallicity, the projected metal density (proportional to the
  mass of metals), and the gas density. 
 The panels extend $250 h^{-1 } \kpc$ in the $x$ and $y$ direction
and $100 h^{-1 } \kpc$ in the $z$-axis.
and are  centered on the quasars (positions shown by crosshairs). The quasars
  have magnitude $m_{B }= -26.3$ (top) and $m_{B }= -26.7$ (bottom) ,
  and the corresponding masses of their host halos are $1.8 \times
  10^{12} h^{-1}\Msun$ and $3.1 \times 10^{12} h^{-1}\Msun$, respectively.}
\label{fig_zpanel}
\end{figure}

\begin{figure}
\centerline{
\psfig{file=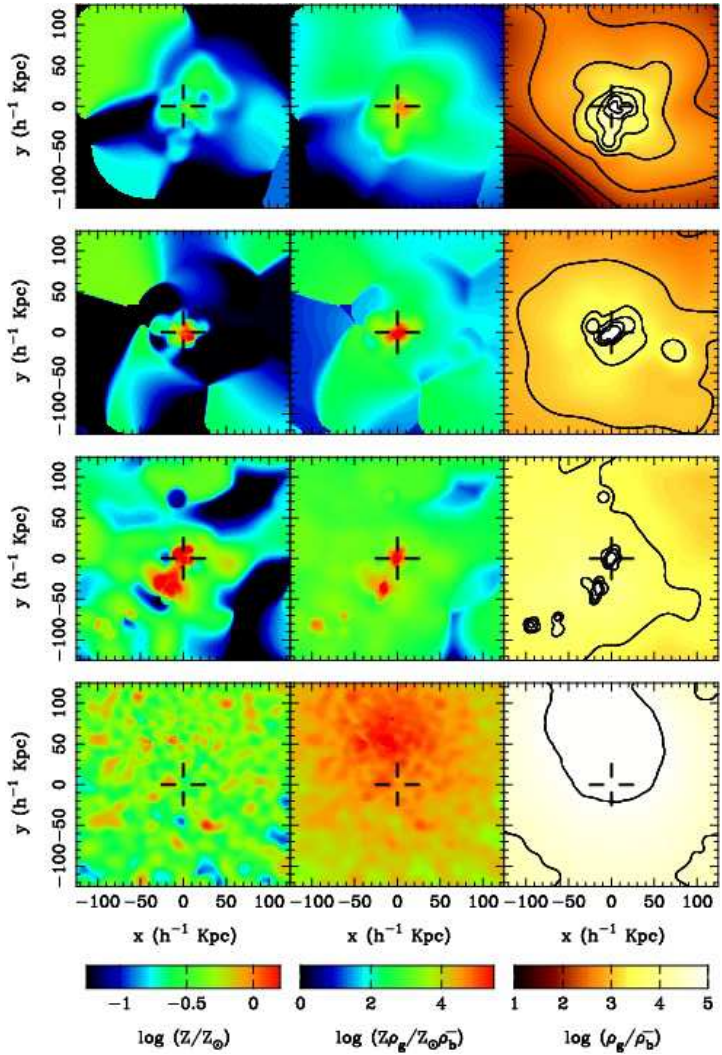,width=9.5truecm,angle=270}
}
\caption{Time evolution of a gas element 
  initially identified in a quasar host at $z=6.5$.  We show
  metallicity (left panels), projected metal density (center panels),
  and projected gas density (right panels), for slices of 
  $100 h^{-1 } \kpc$ width in the $z$-direction and and 250 $ h^{-1 }
  \kpc$ in $x$ and $y$ (same as Fig.~\ref{fig_zpanel}).  The top
  row of panels shows the situation at $z=6.5$, centered on the quasar
  position. At this time the quasar has absolute magnitude $m_{B
  }=-26.2$ and resides in a halo of mass $1.1 \times 10^{12}
  \Msun$. In the succeeding three rows, each panel is centered on the
  position of the particle which was the most bound at $z=6.5$
  (closest to the quasar), but at the later redshifts of $z=4$, $z=2$,
  and $z=0$.}
\label{fig_zsequence}
\end{figure}

\subsection{The quasar hosts}
\begin{table}
\centering
\caption{\label{table_qso}
Quasar host properties.}
\begin{tabular}{cccc}
\hline &&&\\
Redshift  &  $M_{\rm DM}$  & $M_{\rm BH}$ & $SFR$ \\
            & $10^{12} h^{-1} \Msun$   & $10^{9} h^{-1} \Msun$ & $\Msunpyr $\\
\hline &&&\\
 2 &  $3$  & $1.3$& $38$ \\
 5 &  $1.8$  & $0.9 $& $600 $\\
 6.5 &  $1$  & $0.8$ & $270$\\
\hline &&&\\
\end{tabular}
\vspace{-0.8cm}
\tablenotetext{}{\small \\
%\tablecomments{*}
The dark matter, black hole masses and star formation rates for the objects
in Figures~\ref{fig_zpanel} and ~\ref{fig_zsequence}. 
}
\end{table}
Visualizing the morphology of the metal-rich gas in and around quasar
hosts can tell us directly how the metal enrichment occurs in these
high density loci and how uniformly the metals are distributed in
response to star formation and its associated feedback.  In
Figure~\ref{fig_zpanel}, we show a plot of the hosts of two typical
bright quasars (with $B$-band magnitude $m_{\rm b} < -26$), which were
picked randomly from the $z=5$ and the $z=2$ outputs of the G5
simulation. In Table~\ref{table_qso}, we summarize some of the main
properties of these hosts, such as their dark matter, stellar and black
hole masses together with their central star formation rates.  In both
cases, the metallicity close to the center (indicated by the cross on
the figure) is around solar, and the mass of their host halos is
similar and of the order of a few $10^{12} h^{-1}
\Msun$.  The $z=5$ quasar rests in a more concentrated island of metal
rich gas than the $z=2$ case, where the edges of the central
enhancement are more diffuse. It is evident, therefore, that although
the mean metallicity of the ISM is lower at $z=5$, the central regions
do not show significant evolution.  We find that in the central
regions, star formation, and consequently metal enrichment proceeds
rather quickly because of the high densities entailing short star
formation timescales (Eq.~2).  

This plot also shows that star formation occurs in islands of high gas
density that lie nearby in the $z=2$ case, so that the high
metallicity of the gas 50 to 100 $h^{-1}\kpc$ away from the quasar has
likely been generated in situ rather than being blown there by winds.
We note that the star formation rate in the object at $z=5$ is about a
factor 15 higher than in the object at $z=2$, suggesting that the
massive hosts of high redshift quasars are likely to be associated
with more significant star formation activity.

As well as examining at the central region of different bright quasar
hosts active at different redshifts, we can also look at the history
of gas that was once at the center of an early quasar.  In this way,
we can see how the large scale movements of hosts through the evolving
density field impact their metallicity and study the relationship
between the quasar phase and the first major star formation event.  In
Figure~\ref{fig_zsequence}, we follow the position of what was
initially the most tightly bound particle in one of the earliest
bright quasars to be active in the G5 simulation, at $z=6.5$. We can
see that at $z=6.5$, the metallicity near the center of the host is
well below solar, although the mass of the halo is relatively high,
of order $\sim 10^{12} h^{-1}\Msun$. Because in our model the quasar phase is
dependent only on the amount of gas in the halo, it can, at these early
times, be {\it coeval} with the first major event of star
formation in the galaxy.  We note that the star formation rate in the
galaxy measured in the simulation corresponds to $\sim 200 \Msunpyr$
(see Table~\ref{table_qso}).  The metallicity builds up rapidly to
values of around solar in the central region by the $z=4$ timeslice
(and according to Fig.~\ref{fig_zpanel} probably by $z=5$) although
outside $\sim 20 h^{-1}\kpc$ not much has changed. By $z=2$, the
particle has moved into a large scale overdensity, presumably a
forming cluster, and is beginning to be surrounded by metal-rich
gas. At $z=0$, the particle is approximately $75 h^{-1}\kpc$ from the
center of a cluster, which can be clearly seen in the distribution of
metal density. The cluster has no obvious metallicity gradient (at
least in the central portion that can be seen), with no remnant of the
tight knot of high metallicity which was seen in the earlier redshift
panels. The particle which was once close to the center of a quasar
host at $z=6.5$ is now surrounded by gas with $Z\sim0.3$ solar,
typical of the intracluster medium (ICM) observed today.

Although Figure 5 shows only one particular example of the
evolution of the environment of a high redshift quasar, other
quasars, which are bright at these redshifts will have similar
histories. At $z\sim 6-7$ the gas in these high overdensity
regions, where quasars are located, necessarily evolves to
form the central regions of clusters at low redshifts.

\begin{figure}
\centerline{
\psfig{file=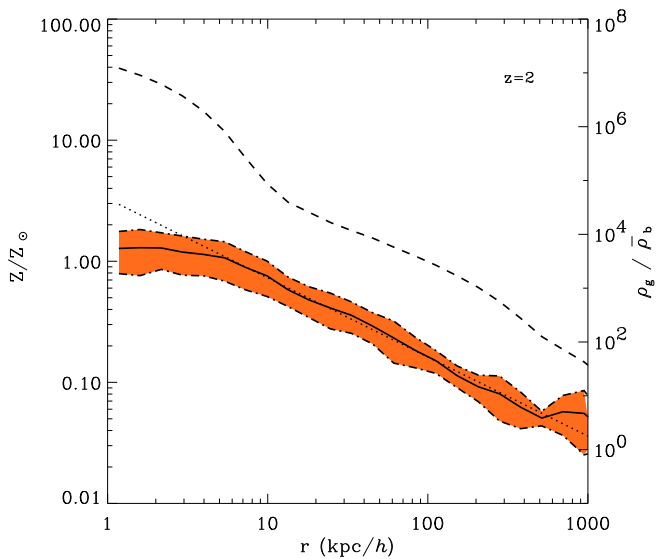,width=9.0truecm}}
\caption{Mean metallicity (left axis) in units of solar metallicity,
  and density profiles in units of the mean baryon density (right
  axis), shown with solid and dashed lines, respectively, for the
  bright quasars ($m_{b} < -26$) at $z=2$ (D5 run). The red area
  represents the standard deviation from the mean metallicity
  profile. Note the similarity between these profiles and those for
  halos of mass $M=2\times 10^{12} h^{-1}\Msun$ shown in
  Figure~\ref{fig_met_m}. The dotted line shows an extrapolation into
  the central region obtained by fitting a power law to the outer
  regions (see text).}
\label{fig_met_qso}       
\end{figure}

\section{Mean quasar metallicity evolution}

\subsection{The circumnuclear metallicity}
In Figure~\ref{fig_met_qso}, we show the mean quasar metallicity
profile and the mean normalized density profile at $z =2$ for the
bright $m_{b} < -26$ quasars. The orange area shows the standard
deviation about the mean metallicity values, and the dotted line gives
a power law fitted to the region outside the core of the profile
(outside $r\sim 10 h^{-1}\kpc$).  Because we have seen that quasars
reside in large mass halos (see also
\S 4) we can only study their properties directly in the larger
simulation boxes of the D5 and G5 simulations. However, according to
the analysis of the numerical convergence in the metallicity profiles
in \S 3, we expect that the true metallicity profiles should exhibit a
power law behavior which continues further into the central regions.

\begin{figure}[t]
\centerline{
\psfig{file=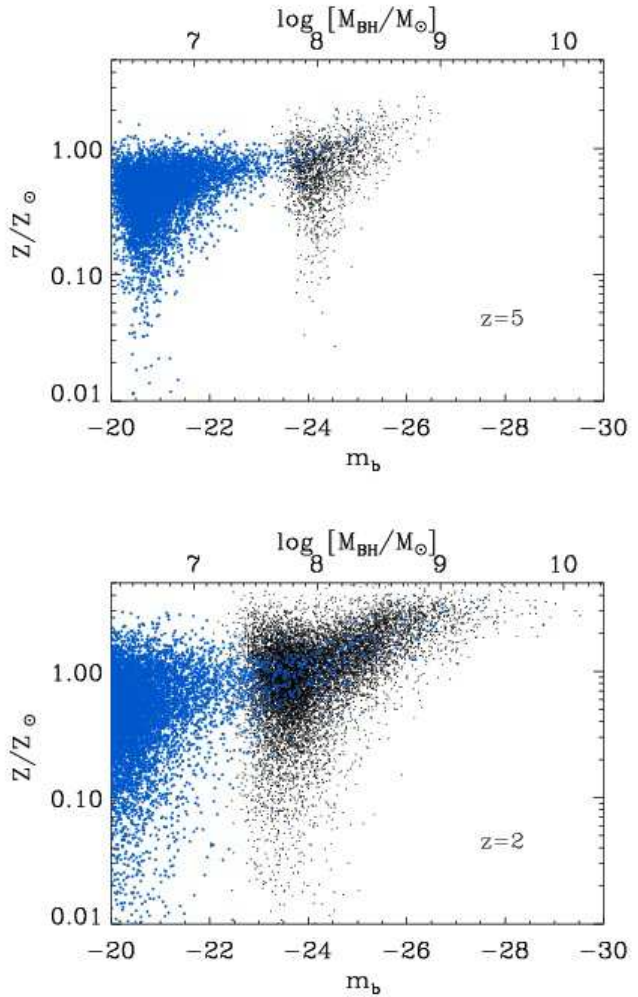,width=10truecm}
}
\caption{Metallicity versus B-magnitude ($m_b$) for all objects in the
  D5 simulation (blue points) and G5 simulation (black) at redshift
  $z=2$ and $z=5$. The comparison between the two different runs shows
  that there is very little dependence of metallicity on $m_{\rm
  b}$. The large scatter seen at low $m_{b}$ is an artifact
  owing to the lack of sufficient resolution close to the mass
  resolution limit of the respective simulation.}
\label{fig_met_qso_mb}       
\end{figure}

It is important to determine the central metallicities from the
simulations because the observed quasar metallicities come exclusively
from the spectroscopic studies of the broad line region and hence
probe the metal enrichment of gas that is in the inner, high density
regions of the galaxy.  Here, we would like to compare the quasar
metallicities obtained from our model to those observed. Given the
relatively strong metallicity gradient that we have found in the
previous sections, and in Figure~\ref{fig_met_qso}, it is necessary
to estimate the metallicity close to the center.

We use the behavior seen in the higher resolution Q5 run to apply a
correction to the metallicities of the quasar hosts in the D5 and G5
simulations. The power law behavior seen in the central parts of the
Q5 galaxies is assumed to apply to the central regions of the D5 and
G5 objects as well. As explained above, we cannot explicitly test the
validity of this well-motivated correction for the most massive
galaxies which make up the hosts of high redshift quasars at
presently observed magnitude limits. This would require yet more
demanding simulations that combine box sizes comparable to that of G5
with a mass resolution as good as that of Q5.  However, we note that
at present, even without our correction, the measured simulation
metallicities represent a robust lower limit, which is above solar,
even at high redshifts.  The metallicities of QSOs fainter than have
yet been observed can be predicted without extrapolation and we
shall do this as well.

For each quasar, we calculate a value for the circumnuclear
metallicities by integrating the metallicity profile mass weighted out
to the radius $R_6$ corresponding to an overdensity of $\rho/\bar{\rho}_{b} =
10^6$ (which typically corresponds to a few \kpc). The metallicities
are then given by
\beq
Z = \frac{\int_0^{R_6} \rho(r) Z(r)
 \,{\rm d}r}{\int_0^{R_6} \rho(r) \, {\rm d}r} .
\label{eq_z}
\eeq
Because we compute
mass weighted metallicity and the slope of the density profile is
relatively shallow in the central regions, our results are fairly insensitive
to the density threshold.

In Figure~\ref{fig_met_qso_mb}, we plot the circumnuclear
metallicities obtain in this way as a function of the quasar $m_{b}$
magnitude at $z=5$ and $z=2$, both for the G5 simulation (black
points) and the D5 simulation (blue points). The plot shows a
moderate trend of increasing metallicity with decreasing quasar
magnitude (although this trend breaks down close to the mass
resolution of the respective simulation, where the scatter in the points
is seen to increase significantly). This trend is broadly similar to
that found by Dietrich et al. (2003b) between metallicity and
continuum luminosity ($L_{\alpha}$) in a sample of high redshift
quasars. It is also consistent with the luminosity-metallicity
relation for quasars (Hamann \& Ferland 1993; 1999, Shemmer \& Netzer
2003). Since we assume universal values for the quasar lifetime and
accretion efficiency, there is a unique black hole mass corresponding to
a given quasar luminosity.  The top axis in
Figure~\ref{fig_met_qso_mb} shows this corresponding black hole mass
for the objects, and hence gives the relation between $Z/Z_{\sun}$ and
$M_{\rm BH}$ inferred from the simulations and the model.

\subsection{Comparison with observations}
\begin{figure}[t]
\centerline{
\vbox{
\psfig{file=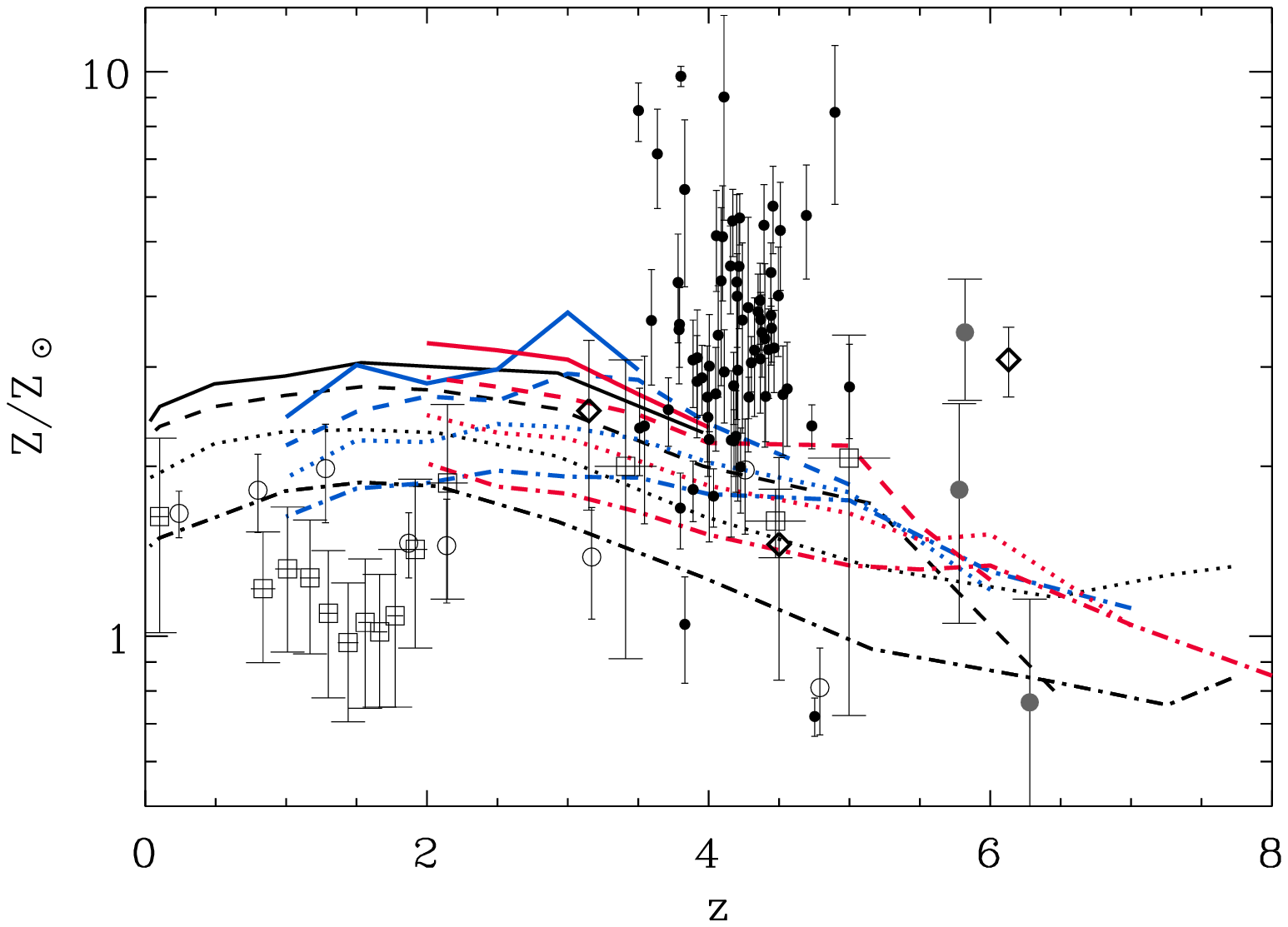,width=9.5truecm}
\psfig{file=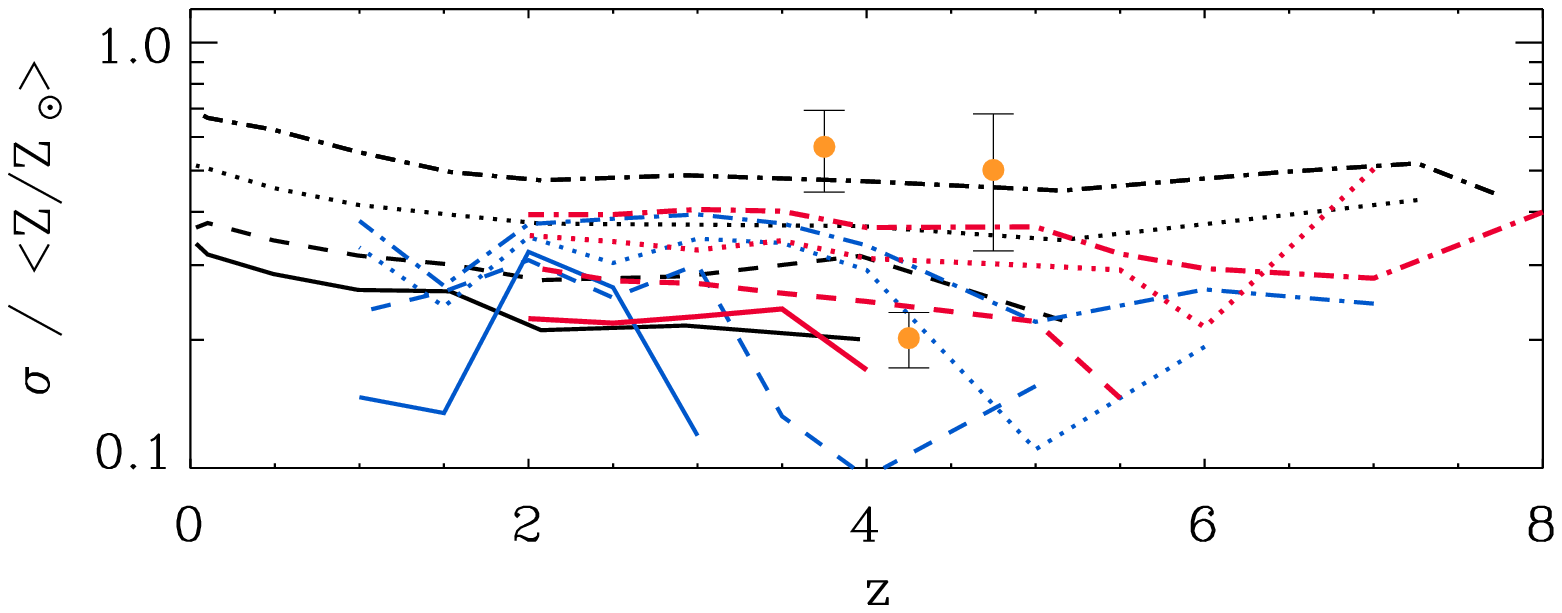,width=9.5truecm}
}}
\caption{Evolution of the mean gas metallicities (top panel) 
  and standard deviations (bottom panel) of QSOs with $m_{b} < -26$ as
  a function of redshift.  The results from the D5 (blue lines), G5
  (black lines) and G6 simulation runs (pink lines) are shown with
  solid, dashed, dotted, dash-dotted lines for quasar lifetimes
  $t_{\rm Q} = 4,\; 2,\; 1,\; 0.5 \times 10^7\,{\rm yr}$,
  respectively. The black solid dots show the measurements of
  individual quasar metallicities by Dietrich et al.~(2003a,b).  The
  solid grey points show the measurement of the flux ratio FeII/MgII
  of 3 SDSS high redshift quasars (Freudling et al.~2003), translated
  into $Z/Z_{\sun} $ by using the factor 2.75 expected for solar
  abundances (Wills, Netzer \& Wills 1985).  The open squares show the
  median and standard deviation (the error bar) from a large sample of
  quasars for which the FeII/MgII flux ratio was measured (Iwamuro et
  al.~2002), the open circles show the same average ratio of FeII/MgII
  measured by Dietrich et al.~(2003c) from a sample of comparable
  luminosity and the open diamonds from a sample of 22 quasar measured
  by Maiolino et al.(2003). The same rescaling factor as above was
  used to obtain the corresponding metallicities.  Metallicities shown
  from these last three sets of observations are more uncertain than
  the Dietrich et al.~(2003a,b) values. In the bottom panel, the
  yellow solid circles show the standard deviation of the quasar
  metallicities in the Dietrich et al.~(2002) sample in three redshift
  bins. The error bars represent Poisson errors.}
\label{fig_Zvsz}
\end{figure}

In Figure~\ref{fig_Zvsz}, we compare observations of quasar
metallicities up to $z \sim 8$ with the mean quasar metallicity
evolution obtained from the simulations for four possible values of
the quasar lifetime, $t_{Q}=4,\;2,\;1,\;{\rm and}\;0.5 \times 10^{7}
\yr$ (solid, dashed, dash-dotted, dotted lines, respectively). The
lines show the predictions for the mean circumnuclear metallicity for
the bright quasars, $m_{b} < -26$, which can be directly compared with
the observations. The black lines are the results from the G5
simulation, the pink lines from the G6
and the blue lines are from the D5 run.

The observational data from Dietrich et al.~(2003a,b) and Freudling et
al.~(2003) represent individual quasar metallicities (solid black and
grey points, respectively). The open squares show the mean metallicity
and standard deviation of a sample of quasars from Iwamuro et
al.~(2002), the open circles are the mean metallicities from a sample
of quasars of comparable luminosities studied by Dietrich et
al.~(2003c) and the diamonds from a sample of 22 quasars studied by
Maiolino et al. (2003). In all these works, the metallicity of the gas
associated with the quasars is estimated from a study of the broad
emission line region (BELR) in the ultraviolet spectral range using
photoionization models (see, e.g., the review by Hamann \& Ferland
1999). The BELR spans a range of distances from a fraction of a pc to
a few pc from the central object, so that it probes the physical
conditions of gas in the circumnuclear region of a quasar.

The solid points in Figure~\ref{fig_Zvsz} represent the chemical
abundances calculated by Dietrich et al.~(2003a,b) from emission line
ratios of several different elements. In particular, nitrogen lines
were related with lines of helium, oxygen and carbon, and the results
from photoionization calculations of Hamann et al.~(2002) were
employed.  The open squares, open circles, open diamonds and grey
points represent measurements of the MgII/FeII ratio from Iwamuro et
al.~(2002), Dietrich et al.~(2003c), Mailino et al.~(2003) and
Freudling et al.~(2003), respectively, which have been converted into
a metallicity using the flux ratio of 2.75 expected for solar
metallicities (derived using photoionization models; see Freudling et
al. 2003; Wills, Netzer \& Wills 1985).  As these last three sets of
data points have been obtained by converting values of MgII/FeII
directly into a metallicity they should be regarded as a far less
reliable indicator of the overall metallicity than the Dietrich et
al.~(2003a,b) data.

The lines in the bottom panel of Figure~\ref{fig_Zvsz} show the
standard deviation, divided by the mean gas metallicity, of the quasar
population in the simulations. For comparison, we have also computed the
standard deviation of the observed quasar metallicities in three
redshift bins (solid yellow points). For this calculation, we subtracted the
measurement errors in quadrature in order to give us an estimate of the
intrinsic scatter in metallicity between quasars. The resulting values
are plotted as symbols in Figure \ref{fig_Zvsz}, with Poisson error
bars. Realistically, these values should probably be taken as upper
limits on the standard deviations because of the uncertainties in the
photoionization modeling used to convert line ratios to metallicities.
However, their magnitude is comparable to the standard deviations from
the simulations.

\begin{figure}[t]
\centerline{
\vbox{
\psfig{file=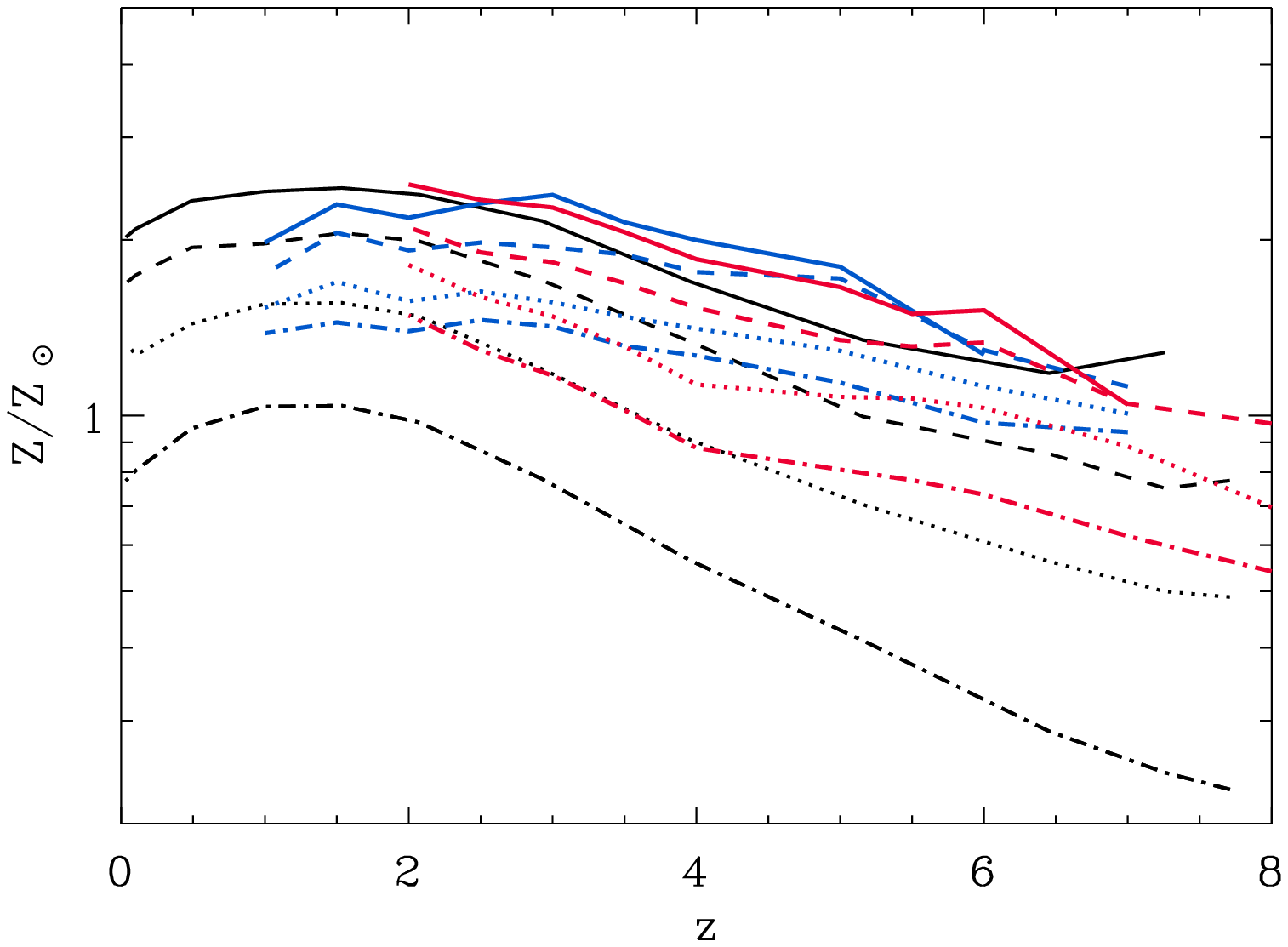,width=9.5truecm}
\psfig{file=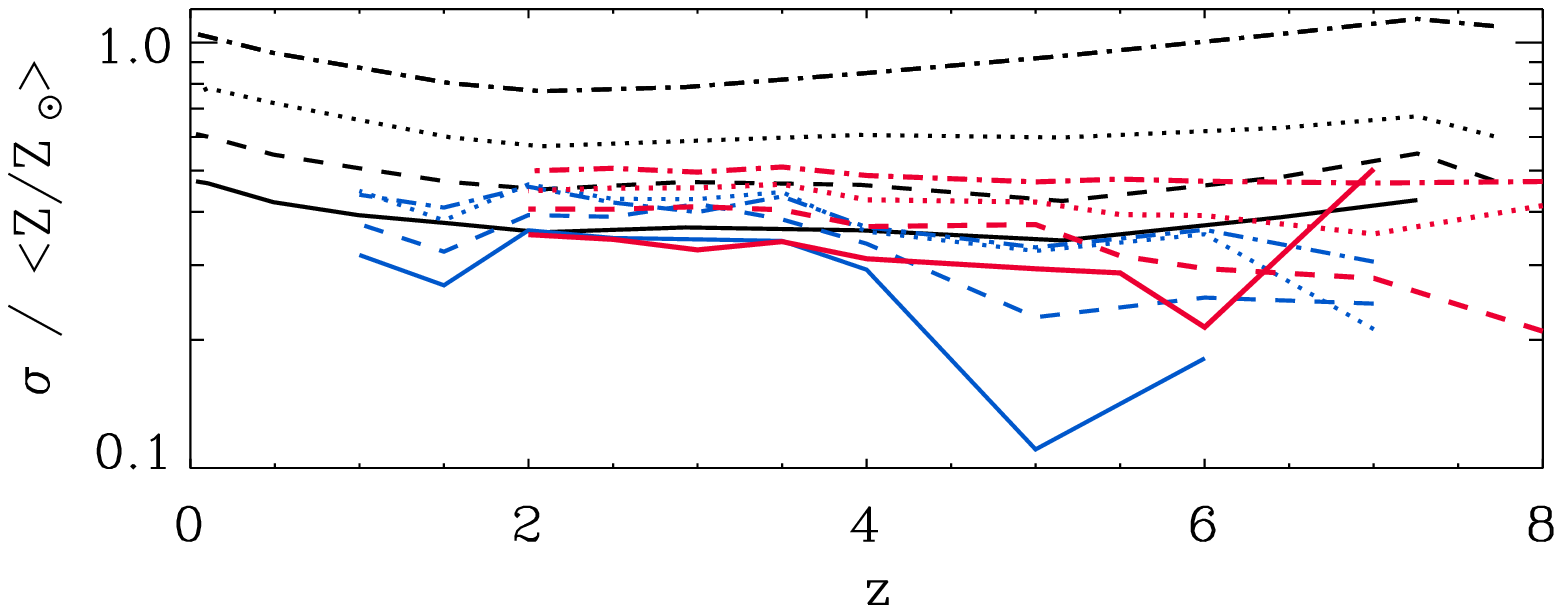,width=9.5truecm}
}}
\caption{Same as Figure \ref{fig_Zvsz},  but including QSOs of lower 
luminosity, down to a cutoff of $m_b < -24.5$.  No data is shown since
a direct comparison with observations is not yet possible.  }
\label{fig_Zvsz_mb}
\end{figure}

The mean metallicity of the gas in the circumnuclear region (as defined
in Eq.~\ref{eq_z}) measured in the simulations reaches values of $Z
\sim 1-3\, Z_{\sun}$, with the highest supersolar values corresponding
to the longest quasar lifetime, $t_{Q} \sim 4 \times 10^7 \yr $. This
trend with $t_{Q}$ is expected because by assuming a longer quasar
lifetime we select very rare high-density peaks to be the hosts of
quasars, whereas a lower timescale implies that quasars that turn on
are found in less biased host galaxies. Note also that the longer
timescales, in our model, are also those that best reproduce the
quasar luminosity function, particularly at $z \approxgt 3$ (Di Matteo
et al.~2003). The corresponding mean gas metallicities of these
massive hosts (see also Fig.~\ref{fig_zpanel}) are comparable to the
range inferred from the observations of the BELRs
(Figure~\ref{fig_Zvsz}).  We emphasize that the supersolar
metallicities in the simulations are associated with the relatively
small fraction of the highest density gas that resides in the
innermost regions of the galaxy (see also
Figs.~\ref{fig_met_res}~and~\ref{fig_met_qso}).  The mean metallicity
of the gas in galaxies as a whole is much lower and reaches at best a
tenth to a few tenths of the solar value (see Fig.~\ref{fig_met_qso}).
It is likely, therefore, that the supersolar chemical abundances
measured in quasars are representative only of the highest density
gas in the centers of galaxies which has undergone star formation on
a short dynamical timescale.

We also note that the mean quasar metallicity shows virtually no
evidence of an evolutionary trend out to redshift $z \sim 5$ (see in
particular the results for the D5 run), which is consistent with all
the current observations (e.g. Dietrich et al.~2003a,b). We predict,
however, a slow decline of the mean metallicity above this redshift,
even though solar-like values are still attained in the high-density
gas up to redshift $z \approxlt 7$.

The differences at $z \approxgt 3$ between the results from the G5 and
D5 runs are due to numerical resolution (as also shown in
Figure~\ref{fig_met_mgas} and Figure~\ref{fig_met_qso_mb}). The higher
resolution runs (like the D5 or Q5) are able to resolve smaller
halos, which are present abundantly at high redshift, better than the
low resolution runs.  This means, for example, that the D5 run has more
star formation, and consequently more metal production at high
redshift than the G5 simulation. On the other hand, because of the
relatively small box size of the D5 run, the largest objects that
contribute to the magnitude limit of $m_{b} < -26$ are very rare at
high-$z$. Making use of the G6 simulation, which has the large box and
higher resolution than the G5 (intermediate between the D5 and G5; see
Table~\ref{table_simul}), we find that for some objects, convergence
is in fact achieved. This is the case for the longest timescales,
which restrict quasar hosts to the most massive objects only. In this
case, the G6 results are consistent with those from the G5 run, 
as we would expect
given that star formation in these most massive halos has been
resolved. The increase in metallicities from redshifts $z \sim 8$ to
$z \sim 5$ is therefore likely to be real in these objects. For
shorter timescales, however, the hosts have typically smaller masses
on average, and convergence is not quite attained (as shown by the
comparison between the G6 and D5 results). Even so, the evolution of
the metallicity for these timescales is expected to be flatter than
for the longer timescales.

It is difficult, therefore, to predict the evolution of the metallicity
beyond redshift $z\sim 5$ in Figure~\ref{fig_Zvsz} for all quasar
timescales reliably. For this reason, and in order to probe the quasar
and halo mass function up to higher redshifts, we include quasars with
lower luminosity down to $m_{b} < -24.5$ in the results shown in
Figure~\ref{fig_Zvsz_mb} (which otherwise has the same format as
Figure ~\ref{fig_Zvsz}). Because of the inclusion of a larger number
of smaller mass halos in this magnitude cut, the difference between
the D5 and G5 runs is increased whereas the G6 and the D5 results do
show convergence for the larger timescales. The general trend of no or
very mild evolution in the gas metallicity is consistent with that in
Figure~\ref{fig_Zvsz}. In addition, we note that shorter quasar
lifetimes seem to imply slightly stronger evolution beyond $z\sim 5$
than the longer lifetimes. Note that the dispersion about the mean is
also larger than in Figure~\ref{fig_Zvsz}. This is expected given the
metallicity-luminosity relation we have found in
Figure~\ref{fig_met_qso_mb}, because we have now considered a larger
range of quasar magnitudes.

By including objects with lower quasar luminosities we are actually
considering those which are less massive and hence more abundant at
high redshifts.  Because of the steep cut-off in the halo mass
function at high redshifts and the association of bright quasars
($m_{b} < -26$) with dark matter masses of the order of $M \sim
10^{12} h^{-1} \Msun$, we simply cannot expect to find these bright objects
beyond $ z \sim 8$.  This implies that in order for observations to
resolve the bulk of the black hole activity beyond these redshifts,
instrument sensitivities need to account for the characteristically
lower quasar luminosities in this regime. If this can be achieved, the
measurement of quasar metallicities provides a potentially useful new
probe for the quasar lifetime and therefore a test of the
characteristic black hole mass function at high redshifts.

\section{Discussion}
We have used cosmological hydrodynamical simulations coupled with a
prescription for black hole activity in galaxies to study the
evolution of the metal enrichment in quasar host galaxies and
explore the relation between star/spheroid formation and black hole
growth/activity.  In the black hole accretion model developed in Di
Matteo et al.~(2003), the black hole fueling rate is regulated by its
interplay with star formation and associated feedback processes in the
gas. This establishes a link between star formation and the quasar
phase in host galaxies, which in turn can broadly reproduce the
observed slope of the relation between black hole mass and stellar
velocity dispersion, as well as the properties of the X-ray and
optical QSO luminosity functions. In this model, the black hole
accretion rate density is found to closely track the star formation
rate density of the simulations, as expected when black hole growth and
fueling are fundamentally linked to the assembly of their host
spheroids.

Using simulations with different resolution, we have measured the
distribution of the gas metallicity and density in quasar host
galaxies (and non-hosts) from $z=8$ to $z=0$. We typically find a
strong radial gradient in the metallicity of the gas in halos, with
the central densest regions of galaxies reaching up to a few times the
solar metallicity (in the most massive objects) while the gas in the
outer regions has a metallicity which is at most around a few tenths
of solar. The more pronounced metal enrichment in the central regions
is a natural consequence of the star formation model in the
simulations which depends on the local gas density and hence occurs on
the shortest timescales in the densest gas ($t_{*} \propto
\rho^{-1/2}$).

We compared the results from the simulations with the observed values
of gas metallicity estimated from the broad emission line region of
bright quasars (e.g. Dietrich et al. 2002, 2003a,b). The simulations
suggest that the super-solar metallicities of quasars deduced from
these studies are representative of the highest density gas that
resides in the innermost regions of galaxies. In this picture, the
very high metallicities probed by the BELR are found only in a very
small fraction of the gas, within the nuclear regions that undergoes
the fastest modes of star formation and chemical enrichment. In other
words, the self-regulated, quiescent mode of star formation
implemented in the simulations can account well for the mildly
super-solar metallicity of the gas seen around bright quasars.

In agreement with observations, the simulations predict little or no
evolution of the mean quasar metallicity in the measured redshift
range of $3.5 < z < 5$ (Dietrich et al. 2003a,b). The mean metallicity
of quasars with $m_{b} < -26$ is predicted to be $Z \sim 2-3\,
Z_{\sun}$ over the redshift range $0 < z < 5$, with a drop by a factor
of 2-3 at earlier times (for $5 < z < 8$).  The high,
solar/super-solar-like metallicities at high redshifts are achieved
quickly in rare and isolated islands of high-density gas residing in
the first halos that form. They thus indicate the presence of massive
star formation already at $z\sim 6-7$ (see
Figs.~\ref{fig_zpanel}~and~\ref{fig_zsequence}).  The scatter between
quasars is likely to give us an idea of the scatter between host
formation times, or between the times of recent major star formation
episodes. Note that the mean gas metallicity of the quasar hosts is
around $~ 0.1-0.5 Z_{\sun}$, typical of massive galaxies
throughout the redshift range.  

We also find a mild dependence of metallicity on quasar luminosity
(and hence black hole mass; Fig.~\ref{fig_met_qso_mb}) which appears
to be consistent with observed trends (e.g. Hamann \& Ferland
1999; Dietrich et al.~2003b; but see also Shemmer \& Netzer 2002).

We note that within the `standard' mode of star formation modeled by
the simulations, the gas metallicity rarely exceeds $Z \sim
4-5\,Z_{\sun}$ whereas some of the most extreme observed values are
reported to be as large as $Z \sim 10\, Z_{\sun}$. As we have 
seen from our convergence tests, the lack of fully resolved star
formation at these high redshifts may still cause us to miss
some metal enrichment. However, the observed
values also bear significant uncertainties owing to the complex
modeling that goes into the measurement of metal abundances from BELR
observations. If such high observed metallicities are indeed present,
 their proper explanation may require higher yields and
hence a top-heavier IMF compared to that in our simulations (which
assume a solar yield). The net yield is related to the IMF $\phi(m)$ by
\beqa
y  = \sum_k \frac{\int_{m_l}^{m_u} m\, p_{k}(m) \phi(m)\, {\rm d}m } 
{1 - \int_{m_l}^{m_u}[m - w_k(m)] \phi(m)\, {\rm d}m},
\eeqa
where $p_{k}(m)$ is the stellar yield of the $k$-th element, and
$w_k(m)$ is the remnant mass of initial mass $m$.  Using $w_k$ and
$p_{k}$ values tabulated by Portinari, Chiosi \& Bressan (1998) and
Marigo (2001) for the mass ranges $6\Msun < m < 120\Msun$ and $0.8
\Msun < m < 6 \Msun$, respectively, and assuming $\phi(m) = {\rm
d}N/{\rm d}m \propto m^{-(1+a)}$, we find that  an IMF-slope of $a
\sim 1.1$
is required to reach a yield approximately twice the solar value,
which is flatter than the Salpeter-IMF with $a=1.35$. For such a
shallower, yet plausible IMF we would then expect (for a simple
closed-box model) that metallicities become larger by about the same
factor of two. The presence of extreme objects like these quasar hosts
with highly super-solar metallicities would then (if confirmed) imply
that such different forms of IMF exist, at least in the central high
density regions of galaxies. It will be interesting to investigate,
with larger observational samples, whether such extreme objects can
only be found at high redshift, and whether there is any evolution of
this extreme population and hence of the typical IMF of massive hosts.

A further constraint on the IMF in the models can be provided by
considering the ionizing background radiation produced by stars in
high redshift galaxies. The observed Lyman-$\alpha$ optical depth in
high redshift quasars constrains the intensity of this radiation. In
Sokasian et al. (2003a,b) the star formation rates from the simulations
were used to predict the intensity of the background radiation. Good
agreement was found with Ly-$\alpha$ observations, assuming an
escape fraction of $20\%$ for the ionizing photons. Changing the IMF
to make it flatter would increase the production rate of ionizing
photons, so that a lower escape fraction would be necessary to match
observations.  With prior information on the escape fraction, a
measurement of the Ly-$\alpha$ optical depth could yield a more
direct constraint on the IMF.

We have shown that the highest values for the mean metallicity, and
hence those which best agree with observations, are obtained for
assumed quasar lifetimes in the range $4\times 10^{7} \approxgt t_{Q}
> 10^{7} \yr$. This is consistent with bright quasars residing in very
massive host galaxies. Shorter quasar lifetimes place quasars also in
less massive hosts where we find that the gas can only achieve
metallicity up to solar like values. This agrees with the
quasar luminosity (and black hole mass)-metallicity relation that we
have shown in Figure~\ref{fig_met_qso_mb}.  We note, coincidentally,
that this range of quasar lifetimes is consistent with constraints
obtained from studies of the reionization of HeII (e.g. Sokasian, 
Abel \& Hernquist 2002, 2003).

Taking into consideration the metallicity inferred from the $\alpha$
elements (the ratio of MgII/FeII), which span a larger range of
redshift, we see that there is an evolution towards lower
metallicities for quasars at $z \approxlt 2$. This may suggest that
below $z \sim 2$ nuclear activity takes place in more common hosts, or
equivalently, that the quasar lifetimes are shorter than in the high
redshift quasars (but note that these measurements have more
significant uncertainties when used to infer global metallicities than
those obtained from N, O, and C, and maybe also include objects
spanning a larger range of luminosities).

Using the simulations, we can directly look at the predicted
properties of quasar host galaxies at high redshifts. In our model,
bright rare quasars at $z \approxgt 5$ reside typically in massive
dark matter potentials, with $M_{\rm DM} \sim$ a few $10^{12} h^{-1}\Msun$,
characterized by high star formation rates of the order of $10-100
\Msunpyr$. Therefore, quasars reside in the rare, most massive halos
which are already forming stars vigorously at these times. 
Most of the black hole growth occurrs
simultaneously with the processes that dump matter into the central
regions of galaxies and induce substantial star formation. In
particular, under the assumption that the black hole mass is
proportional to the mass of gas, we find typical black hole masses for
these objects of order $M_{\rm BH} \sim 10^{9}
\Msun$. We find that the quasar phase at high redshift follows or is
roughly co-eval with the major star formation event in these massive
objects (see Figs.~\ref{fig_zpanel}~and~\ref{fig_zsequence}).  These
results are consistent with sub-mm and IR observations of high
redshift quasars which imply large luminosities in these bands and
hence gas rich, strongly star forming hosts (Omont et al.~2001;
Archibald et al 2001; Pridley et al.~2003).

The results we have discussed here further strengthen the evidence for
a strong relation between galaxy formation, star formation and black
hole activity, and hence underline the importance of quasar studies
for understanding high-redshift star formation and early galaxy
evolution.  Using the simulations and the measured metal abundances,
we have shown that it is possible to construct self-consistent models
for the locations where massive galaxies are being assembled,
vigorously forming stars and building central black holes.  As more
precise measurements of quasar metallicities become available, it may
therefore become possible to use these models to much better constrain
the parameters which govern the stellar mass function in the redshift
range where the large quasar hosts were forming.  It should be within
the capabilities of forthcoming millimeter array telescopes (like
ALMA) and the James Webb space Telescope (JWST) to observe quasars in
this fashion up to redshift $z\sim 8$ and beyond.

\acknowledgements
TDM thanks Sofia Cora for useful discussions and for providing the
stellar yield tables and Matthias Dietrich for discussions
and the tabulated quasar metallicities.
This work was supported in part by NSF grants ACI
96-19019, AST 98-02568, AST 99-00877, and AST 00-71019 and NASA ATP
grant NAG5-12140.  The simulations were performed at the Center for
Parallel Astrophysical Computing at the Harvard-Smithsonian Center for
Astrophysics.


\begin{references}
  
  \vspace{1cm} 


\reference{}Aguirre, A., Hernquist, L., Schaye, J., Weinberg, D., Katz, N.,
Gardner, J., 2001a, ApJ, 560, 599
\reference{}Aguirre, A., Hernquist, L., Schaye, J., Katz, N., Weinberg, D.,
Gardner, J., 2001b, ApJ, 561, 521
\reference{} Archibald, E.~N., Dunlop, J.~S., Hughes,
  D.~H., Rawlings, S., Eales, S.~A., \& Ivison, R.~J.\ 2001, \mnras,
  323, 417
\reference{} Carilli, C. L., Bertoldi, F., Menten, K. M.,
 Rupen, M. P., Kreysa, E., Fan, X.,
 Strauss, M. A., Schneider, D. P.,
 Bertarini, A., Yun, M. S., Zylka, R., 2000, ApJ, 533, L13
\reference{}Carilli, C. L., Kohno, K.,
 Kawabe, R., Ohta, K., Henkel, C.,
 Menten, K. M., Yun, M. S., Petric, A.,
 Tutui, Y., 2002, ApJ, 569, 605.
\reference{}Ciotti L., van Albada T.S., 2001, ApJ, 552, L13
\reference{} Cox, P., Omont, A., Djorgovski, S. G.,
 Bertoldi, F., Pety, J., Carilli, C. L.,
 Isaak, K. G., Beelen, A., McMahon, R. G.,
 Castro, S., 2002,  A \& A, 387, 406
\reference{}Croft R. A. C., Di Matteo T., Dav\'e R., Hernquist L., 
Katz N., Fardal M. A., Weinberg D.H., 2001, ApJ, 557, 67
\reference{}Dav\'e, R., Hernquist, L., Katz, N., Weinberg, D.H., 1999,
ApJ, 511, 521
\reference{} Dietrich, M., Appenzeller, I., Wagner, S. J., Gässler, W., 
Hafner, R., Hess, H.-J., Hummel, W., Muschielok, B., Nicklas, H.,
Rupprecht, G., Seifert, W., Stahl, O., Szeifert, T., 
Tarantik, K., 1999, A\&A, 352, L1
\reference{} Dietrich M., Appenzeller I., Vestergaard M., Wagner S.J.,
  2002, ApJ, 564, 581
\reference{} Dietrich M., Appenzeller I., Hamann F, Heidt J., Jager K., Vestergaard M. \& Wagner S.J., 2003a, A\&A, 398, 891
\reference{} Dietrich M., Hamann F., Shields J.C., Costantin A., Heidt J., Jager K., Vestergaard M.,
Wagner S.J., 2003b, ApJ, 589, 722
\reference{} Dietrich M., Hamann F., Appenzeller I., Vestergaard M.,
  2003c, ApJ, 596, in press 
\reference{} Di Matteo T., Croft R.A.C., Springel V., Hernquist L., 
2003, ApJ, 593, 56
\reference{}Efstathiou, G \& Rees, M. J., 1988, \mnras, 230, 5
\reference{}Ferrarese L. \& Merritt D., 2000, ApJ, 539, L9
\reference{}Freudling W., Corbin M.R., Korista K.T., 2003, ApJ, in press
[astro-ph/0303424]
\reference{}Gebhardt K. et al., 2000, ApJ, 539, L13
\reference{} Granato G.L., De Zotti G., Silva L., Bressan A., Danese L., 
2003, submitted [astro-ph/0307202]
\reference{} Haiman Z., Ciotti L., Ostriker J.P., 2003, ApJ, submitted
[astro-ph/0304129]
\reference{} Hamann, F., Ferland, G., 1993, ApJ, 418, 11
\reference{} Hamann, F., Ferland, G., 1999, ARA\&A, 37, 487
\reference{} Hamann, F., 
Korista, K. T., Ferland, G. J.,
 Warner, C., Baldwin, J., 2002, ApJ 564, 592
\reference{}Hernquist L., 1993, ApJ, 404, 717
\reference{}Hernquist L., Springel V., 2003, MNRAS, 341, 125
\reference{} Iwamuro, F., Motohara K., Maihara, T., Kimura M., Yoshi Y., Doi M., 2002, ApJ, 565, 63 
\reference{}Katz, N., Weinberg, D.H., Hernquist, L., 1996, ApJS, 105, 19
\reference{}Kauffmann G., Haehnelt M., 2000, MNRAS, 311, 576 
\reference{}Kennicutt, R.C.Jr., 1998, ARA\&A, 36, 189
\reference{}Kogut, A. et al., 2003, ApJ, submitted [astro-ph/0302213]
\reference{}Kormendy J., Gebhardt K., 2001, In Wheeler J.C. Martel
H. eds, AIP Conf. Proc. Vol. 586, 20th Texas Symposium On Relativistic
Astrophysics. Am. Inst. Phys., New York, p.363
\reference{}Magorrian et al., 1998, AJ, 104, 372
\reference{}Maiolino R., Juarez Y., Nagar N.M., Oliva E., 2003, ApJ,
in press, [astro-ph/0307264]
\reference{}Marigo P., 2001, A\&A, 370, 194
\reference{}Merritt D., Ferrarese, L., 2001, MNRAS, 320, L30
\reference{} Monaco, P., Salucci P., \& Danese L., 2000, MNRAS, 311, 279
\reference{} Omont, A., Cox, P., 
Bertoldi, F., McMahon, R.~G., Carilli, C., \& Isaak, K.~G.\ 2001, A\&A, 
374, 371 
\reference{} Portinari L., Chiosi C., Bressan A., 1998, A\&A 334, 505
\reference{} Pridley R.S., Isaak K.G., McMahon R.G., Robson E.I.,
Pearson C.P., 2003, MNRAS, submitted [astro-ph/0308132]
\reference{} Shemmer O. \& Netzer H., 2002, ApJ, 567, L19  
\reference{}Sokasian, A., Abel, T., \& Hernquist, L.\ 2002, MNRAS, 332, 601
\reference{}Sokasian, A., Abel, T., \& Hernquist, L.\ 2003, MNRAS, 340, 473 
\reference{}Sokasian, A., Abel, T., Hernquist, L., \&
Springel, V.\ 2003a, MNRAS, 344, 607 
\reference{}Sokasian, A., Abel, T., Hernquist, L., \&
Springel, V.\ 2003b, MNRAS, submitted [astro-ph/0307451]
\reference{}Springel V., Hernquist L., 2002, MNRAS, 333, 649
\reference{}Springel V., Hernquist L., 2003a, MNRAS, 339, 289
\reference{}Springel V., Hernquist L., 2003b, MNRAS, 339, 312
\reference{}Springel V., Yoshida, N., White, S.D.M., 2001, NewA, 6, 79
\reference{}Tremaine S. et al. 2002, ApJ, 574, 740
\reference{}Wills B., Netzer, H., Wills, D., 1985, ApJ, 288, 94 
\reference{}Wyithe J.S., Loeb A., 2002, ApJ, 581, 886
\reference{}Yoshida, N.,  Sokasian, A., Hernquist, L., \& Springel, V.,
 2003a, ApJ, 591, L1
\reference{}Yoshida, N.,  Sokasian, A., Hernquist, L., \& Springel, V.,
 2003b, ApJ, in press [astro-ph/0305517]

\end{references}
\end{document}